\documentclass[manuscript]{aastex}

\slugcomment{}

\newcounter{subfigure}

\shorttitle{Infrared Diffuse Interstellar Bands}
\shortauthors{Rawlings et al.}

\begin{document}


\title{A High-Resolution Study of Near-Infrared Diffuse Interstellar Bands}


\author{M. G. Rawlings}
\affil{National Radio Astronomy Observatory,
520 Edgemont Rd,
Charlottesville, VA 22903, USA.}
\email{mrawling@nrao.edu}

\author{A. J. Adamson}
\affil{Gemini Observatory,
670 N. A`ohoku Place,
University Park,
Hilo, HI 96720,
U.S.A. }
\email{aadamson@gemini.edu}

\and

\author{T. H. Kerr}
\affil{Joint Astronomy Centre,
660 N. A`ohoku Place,
University Park,
Hilo, HI 96720,
U.S.A. }
\email{t.kerr@jach.hawaii.edu}




\begin{abstract}
We present high-resolution echelle spectroscopic observations of the two near-infrared (NIR) Diffuse Interstellar Bands (DIBs) at 13175\AA\ and 11797.5\AA. The DIBs have been observed in a number of diffuse interstellar medium sightlines that exhibit a wide range of visual extinctions. Band profiles are similar to those seen in narrow DIBs, clearly asymmetric and can be closely fitted in most cases using two simple Gaussian components. Gaussian fits were generally found to be more successful than fits based on a multiple-cloud model using a template DIB profile. For a sample of 9 objects in which both bands are observed, the strength of both NIR DIBs generally increases with {\it A(V)}, and we report a correlation between the two observed bands over a large {\it A(V)} range and widely-separated lines of sight. The strength of the two bands is also compared against those of two visual DIBs and the diffuse ISM aliphatic dust absorption feature at 3.4$\mu$m previously detected in the same sightlines. We find that the NIR DIBs do not exhibit notable (anti)correlations with either. Implications of these observations on possible DIB carrier species are discussed.
\end{abstract}





\section{INTRODUCTION}

\subsection{DIBs: History and Key Characteristics}

Diffuse Interstellar Bands (DIBs) are ubiquitous absorption features seen in the spectra of bright background sources (e.g. stars) that arise from the presence of foreground diffuse interstellar materials. Since the first DIB detections (\citealt{heger1922}; \citealt{merrill1934}), the identification of their carriers has become one of the longest-standing problems in astrophysics. To date, several hundred DIBs have been detected at visual wavelengths (see e.g. \citealt{hobbs2008}, \citealt{hobbs2009}). Although the DIB carriers are now widely held to be carbonaceous molecules/ions, a specific identification remains elusive. The nature of the DIB carrier(s) has implications for interstellar chemical processes, cosmic elemental abundances and gas-grain interactions.

One of the most significant DIB discoveries in the last few decades was that many of the bands exhibited fine structure.  The first signs of this were found by \citet{herbig1982}, with subsequent observations characterizing DIB asymmetry in detail \citep{westerlund1988a, westerlund1988b, krelowski1988}.  Following this, studies at high and ultra-high resolutions of some of the stronger DIBs, such as those at 5797\AA~(17250 cm$^{-1}$) and 6614\AA\ (15119 cm$^{-1}$; \citealt{sarre1995, ehrenfreund1996, kerr1996, jenniskens1996, krelowski1997, kerr1998}) have shown that multiply-peaked substructures can occur in narrow DIBs.  \citet{sarre1995} noted that the wavelength separations between the three primary subfeatures of $\lambda\lambda$5797 and 6614 were similar, but the presence of other subfeatures complicated the profiles beyond those of simple gas-phase PQR profiles.  Subsequently, \citet{kerr1998} used resolving powers of up to 600000 to detect near-identical ultrafine structure in $\lambda$5797 towards $\mu$ Sgr, $\zeta$ Per and $\zeta$ Oph.  They were able to identify {\it seven} subfeatures within the one narrow band, all within 0.5\AA\ of the central wavelength. Other studies of DIB substructure have attempted to use them as characteristics for possible DIB groupings \citep{galazutdinov2008a, galazutdinov2008b}.

In addition to the numerous visual-wavelength DIBs that have been observed, two DIB detections were reported in the near-infrared by \citet{joblin1990} near 11797.5\AA\ (8476 cm$^{-1}$) and 13175\AA\ (7590 cm$^{-1}$). A subsequent non-detection of the 13175\AA\ band toward the comparatively dense Taurus region \citep{adamson1994} supported its identification as a DIB. More recently, an additional thirteen features were also identified as NIR DIBs \citep{geballe2011}. All of these bands occur at wavelengths notably longer than those of all of the other known DIBs, and this may have implications for the identification of their carriers. There are few neutral molecules known to have electronic transitions at such low frequencies (in contrast to the visual band, in which many molecules have transitions), suggesting that the carriers have a very small spacing between their ground electronic state and the first excited state.  However, a number of PAH ions are known to have NIR electronic transitions \citep{mattioda2005}. Further study of these two bands could therefore potentially lead to additional clues to the nature of the DIB carriers, assuming their carriers also give rise to other bands. Despite their properties, however, comparatively little follow-up work has been conducted on the NIR DIBs since their discovery.

\subsection{Scientific Aims, and Objects Selected for Observation}

Although echelle spectroscopy resolutions currently reachable in the NIR are lower than those in the visual, they can still be very useful for molecular spectroscopy. This is because for a given resolving power, such observations are as effective when resolving directly comparable molecular substructure at longer wavelengths. For example, if two bands are identified at 0.65$\mu$m and 1.3$\mu$m as having similar widths in  frequency (and hence energy) terms, the band with twice the central wavelength should also have twice the width in wavelength terms.

We obtained high-resolution UKIRT/CGS4 echelle spectra of the 11797.5\AA\ and 13175\AA\ DIBs for a range of diffuse interstellar medium (DISM) targets.

In selecting the observational targets for this work, the specific aims were as follows:

\begin{enumerate}
\item To confirm whether or not the NIR features are indeed DIBs;
\item To attempt to identify and resolve any NIR DIB asymmetry and/or substructure that might be present;
\item To enable the deconvolution of the possible effects of multiple cloudlets in high-extinction sightlines;
\item To enable the comparison of the NIR and visual DIB strengths for a range of values of {\it A(V)}.
\end{enumerate}

To achieve these objectives, we obtained CGS4 echelle spectroscopy of the 11797.5\AA\ and 13175\AA\ DIBs in sightlines with a range of DISM extinction levels and bright, early-type background field stars with comparatively featureless intrinsic spectra. The source list selected was as follows:

\begin{enumerate}
\item {\bf Several known high-extinction diffuse medium sightlines towards both low and high Galactic longitudes (the Galactic Center and Cygnus respectively):} These targets include the classic high-extinction sightline towards the hypergiant Cyg OB2 No. 12, and several of the early-type Stephenson star sightlines \citep{rawlings2003} that have already been identified as exhibiting both high diffuse ISM extinction and strong DIBs in the visual band.
\item {\bf $\mu$ Sgr:} This bright target was chosen as a ``single cloud'' sightline, and was selected as a control observation to obtain NIR band profiles without the ambiguity arising from the possible contributions of multiple line-of-sight cloudlets. It is also already known to exhibit well-defined substructure in the 5797 and 6614\AA\ DIBs \citep{kerr1996}.
\item {\bf HD 204827:} Extensive DIB spectroscopy at visual wavelengths exists for this spectroscopic binary source \citep{hobbs2008}. Observations of this target were performed at several orbital epochs, in order to further confirm the DIB nature of the features.
\end{enumerate}

\section{OBSERVATIONS AND DATA REDUCTION}

The targets were observed during the period of nights   2008 August $9 - 21$ UT from the United Kingdom Infrared Telescope (UKIRT), Hawaii, using the echelle grating in CGS4. Weather was clear for almost the entire period. A 1-pixel width slit (0.6'') of length 91.5'' was used. Slit position angles for each target were chosen in advance so as to avoid other field targets. 2 $\times$ 2  sampling was used. Observations were acquired in non-destructive readout mode using a standard ABBA nod of 16 pixels along the slit. Observations of the 13175\AA\ DIB were conducted at a specified central wavelength of 1.3175 $\mu$m, with a wavelength coverage of 1.3135 -- 1.3215 $\mu$m. Observations of the 11797.5\AA\ DIB were conducted at a specified central wavelength of 1.17975 $\mu$m, with a wavelength coverage of 1.1762 Ð 1.1833 $\mu$m. The effective resolution was $\sim$37000. Integration times were specified and adjusted on a source-by-source basis in order to avoid saturation, with the aim of achieving a continuum signal-to-noise ration (s/n) of at least 100 on all of the targets (except for $\mu$ Sgr) in order to obtain a $\sim5\sigma$ or better detection of the DIBs. A target continuum s/n of $\sim500$ was assigned to $\mu$ Sgr, since as the token single-cloud sightline, its DIB was expected to be comparatively weak. Accompanying airmass-matched Bright Standard (BS) stars were observed to provide flux calibration. These were typically late-B-type giant stars, in order to ensure relatively featureless photospheric spectra at the observed wavelengths. Flat field observations and corrections were applied automatically for each image frame. observed. For the 11797.5\AA\ observations, arc line spectra of krypton and xenon were also observed prior to each target. For the 13175\AA\ observations, arc line spectra of krypton and argon were observed. In practice, however, it was found that most of the arc lines at these wavelengths were too weak for reliable wavelength calibration. Wavelength calibration for both observed spectral ranges was therefore instead performed using telluric absorption features following their identification against standard line lists (\citealt{wallace2000} and references therein).

The data were reduced using the Starlink software suite, specifically Figaro, KAPPA and GAIA. The basic data reduction process involved a number of steps. Firstly, the positive and negative stellar spectra were extracted from each of the co-added ABBA groups. Any bad pixels arising due to cosmic ray strikes were interpolated out. The extracted 1D spectra were registered and coadded. These spectra were then wavelength-calibrated against telluric line lists (see above). This same process was also used to extract wavelength calibrated, normalized spectra for the observed standard stars. Science target spectra were then registered with suitable standards, and the ratios of suitable science target/standard star pairs were taken to cancel telluric absorption. For the cases in which only poor airmass matches to the science targets were available, other observed standard star spectra were tried, and/or suitable logarithmic rescaling of the standard star spectrum used \citep{rawlings2003} was performed to optimize telluric cancellation. The resultant spectra then normalized by ratioing them with low-order polynomial continuum fits.




\section{RESULTS}

\subsection{Detections and Profiles}

The resultant rectified spectra are shown in Figure 1 (\ref{fig1a}, \ref{fig1b} and \ref{fig1c}), along with their Gaussian fits (see the next subsection for more details). For all the observed sources except two (StRS 371 and HD 204827, for which no 11797.5\AA\ were obtained), the left- and right-hand panels of Figure 1 show the 11797.5\AA\ and 13175\AA\ DIBs respectively. Both of the NIR DIBs are clearly detected in all observed sources. The s/n of the HD 204827 spectra were much lower than expected. In light of this, a representative spectrum from a single epoch is presented as part of Figure 1(c), in which the stronger DIB is clearly detected. However, a detailed deconvolution of the DIB profile from the stellar spectral line movement (due to source binarity) was not possible.

Structure seen in the Figure 1 spectra near 1.317 $\mu$m is generally attributable to imperfect cancellation of the telluric standards. The two notable exceptions are StRS 164 and StRS 217, for which absorption is clearly seen in the unratioed science target spectra. On the basis of H$_3^+$ spectroscopy, this absorption is not part of the 13175\AA\ DIB, and remains unidentified at this stage. In no case does this 1.317 $\mu$m structure impact on the profile of the DIB itself, and telluric correction in the vicinity of the DIB is generally very good.

Figures 2 and 3 show the two diffuse bands separated into groups based on the profile shape per band.  In both cases, a ÒbaselineÓ (majority) set is apparent and other profiles are shown along with that baseline. Cyg OB2 No. 12 is in the baseline group in both cases, and as it has the highest s/n, it is used as the archetype against which the other groups are compared. The differences between these these groups appear to be real and are not a function of s/n or telluric cancellation.

\subsection{Gaussian Characterization of the DIBs}

The DIB profiles appear to be relatively simple, but significantly asymmetric. They are quite well fitted in all cases by two or fewer Gaussian components. The two observed DIBs, however, could be fitted well by up to two Gaussians in all cases. The results of these fits are shown in Figure 1 and Tables 2 and 3. Similar profile shapes have previously been seen in multiple narrow visual-wavelength DIBs (e.g. \citealt{krelowski1997}). Towards $\mu$ Sgr, the same behavior is observed in the 13175\AA\ DIB profile, but not in the (relatively weak) 11797.5\AA\ DIB (Figure 1(a), second row panels). Neither of the observed DIBs show more complex substructure at the spectral resolution used here ($\sim$37000).

Uncertainties on the overall fits for each source (listed in Tables 2 and 3) were determined via a series of several steps. A first approximation of the error on each of the two Gaussians was obtained by estimating the r.m.s. noise on the spectral continuum and adding/subtracting that from the depth of the Gaussian. The combined  "added" and "subtracted" values were taken as the as the absolute upper and lower plausible limits. The final estimated compound errors were then taken to be the half of this total possible range divided by $\sqrt 2$ (to allow for the non-additive, random nature of the uncertainty).

Figure 4 shows how the fitted strengths of the two bands vary with {\it A(V)}. The upper plot shows our data for the 11797.5\AA\ DIB, and the lower plot shows our data for the 13175\AA\ DIB, plotted together with the data from \citet{joblin1990} and the upper limits obtained by \citet{adamson1994}. Based on the combined dataset of Figure 4, the 13175\AA\ DIB generally appears to exhibit a linear correlation with {\it A(V)}, although two of our Cygnus sources in the lower right of the plot (StRS 344 and 368) appear to have weaker DIBs than expected for their high {\it A(V)} values (13.1 and 14 respectively). A similar, weaker general trend is seen for the 11797.5\AA\ DIB. We return to this point in Section 4.

Figure 5 illustrates the relative fitted strengths of the two DIBs. The upper panel of Figure 5 shows the measured DIB strengths, which are clearly strongly correlated. The lower panel shows the DIB strengths normalized by {\it A(V)}. For this plot, the Pearson correlation coefficient for the sources with measurements of both DIBs was 0.94. However, previous observations at visual wavelengths indicate that DIBs toward $\mu$ Sgr are unusually strong per unit {\it A(V)}  \citep{kerr1996}, and hence may not be representative of DIB strengths elsewhere. Conversely, many of the visual band DIBs measured toward HD 204827 are relatively weak per unit {\it A(V)} \citep{hobbs2008}. This is consistent with the weak 13175\AA\ band seen in Figure 4 and Table 3, and our lack of a clear detection of the 11797.5\AA\ DIB towards this target.

A recalculation of the Pearson correlation coefficient for the dataset with the $\mu$ Sgr data point excluded yields a value of 0.56. As with the visual DIBs toward these stars \citep{rawlings2003}, the low- and high-longitude lines of sight are statistically indistinguishable. A correlation in this plot would suggest some physical relationship between the carriers of the two bands, such as similarly favorable formation conditions and/or similar carrier species. We also note, however, that errors in the determination of {\it A(V)} may affect the correlation (or lack of one) in the lower panel of Figure 5 (see Table 3 of \citealt{rawlings2000} for details). The systematic error estimates on the spectroscopically-determined {\it A(V)} values quoted there are $\pm 0.1 - 0.3^m$, and for internal self-consistency, we have adopted those estimates since photometric determinations are not available for the majority of the sources observed here. For the four sources listed in this paper for which photometric {\it A(V)} values are given in \citet{rawlings2000}, the photometric {\it A(V)} values are lower, but still all within $33\%$ of the spectroscopic values used here.

\section{DISCUSSION}

\subsection{The Bands as DIBs}

There seems little doubt that the two bands observed toward our chosen sightlines are DIBs. Since our HD 204827 data were not of sufficient quality to allow the deconvolution of the NIR DIBs from the moving stellar lines in the binary star spectra, they are not considered further here. In all other respects, however, the features do appear to behave as DIBs, and the majority of spectral features already identified as DIBs elsewhere in the literature have also not been explicitly examined via binary star spectra. Their profile shapes, comparative narrowness and relatively minor variations in wavelength are typical of many DIBs seen at other wavelengths. Furthermore, a detection of the 13175\AA\ DIB was also reported towards the source qF362 by \citet{geballe2011}, the spectrum of which was also found to exhibit a number of additional strong NIR DIBs.

\subsection{Relative Strengths and Profiles of the NIR DIBs}

As can be seen in Figure 1, the 13175\AA\ DIB is markedly stronger in equivalent width than the 11797.5\AA\ DIB towards all of the sightlines observed. This is consistent with previous observations of these two DIBs elsewhere, and suggests that the  $\lambda$13175 carrier has a higher intrinsic efficiency and/or is more abundant. There appears to be no strong evidence for subfeatures of the type seen in some of the narrow visual-wavelength DIBs, at least at the resolutions attainable with UKIRT/CGS4 (R$\sim$37000). 

The overall profiles of the DIBs appear to be at least superficially similar to, or even typical of, other narrow DIBs observed at shorter wavelengths. Both DIBs frequently exhibit marked asymmetry due to an enhancement of the red wing. The left panels of Figure 6 clearly show similarities between the observed NIR DIB profiles toward Cyg OB2 No. 12 and averaged visual DIB profiles measured towards other sightlines \citep{krelowski1997}. However, the profiles vary significantly between lines of sight; this is illustrated in the two rows of panels in Figure 6. This is sufficiently common in other DIBs that \citet{mccall2010} cautioned that the presence of such extended wings can complicate the accurate characterization of DIB-DIB correlations. Several attempts have been made in the past to explain such asymmetry in visual DIBs. For example, a number of possible mechanisms have been proposed that broaden PR or PQR branch electronic transition spectra to yield the observed DIB profiles. In this latter picture, the breadth of the DIBs is attributed to some kind of internal conversion process, in which the upper electronic state has a finite lifetime before decay, resulting in a lifetime broadening of the band. No molecule has a rotational contour that closely resembles a Gaussian or a Lorentzian profile, and so if the broadening is not sufficiently strong, some observable residual structure from (e.g.) the P, Q, and R branches would remain, or at least the contour of that structure.

The width of the two DIBs varies somewhat between sightlines, and some sightlines also appear to exhibit evidence of blue wings. We therefore consider here the possibility that this is reflects velocity structure smearing due to the motions of multiple clouds in the line of sight. The general profile asymmetry is still comparatively subtle at the current spectral resolution, and recurs along multiple sightlines, including the 13175\AA\ DIB towards $\mu$ Sgr. It therefore seems unlikely that the observed red wing structure is purely attributable to velocity structure. 

To constrain the ability of velocity structure to match the observed profiles, we attempted to fit them with template spectra based on the single-cloud line of sight toward $\mu$ Sgr. The templates adopted were the idealized two-Gaussian fits for this object shown in Figure 1.

Preserving the band shape reflected in those fits and allowing only the overall central wavelength and depth to vary, each spectrum was fitted allowing the presence of up to three Ò$\mu$ SgrÓ clouds, guided by spectroscopy of the H$_3^+$ doublet near 3.668 $\mu$m \citep{mccall1998}, providing up to six fitting parameters for these particular sources (T. R. Geballe, private communication).

For the 11797.5\AA\ DIB, the presence of a blue wing in the ``intrinsic'' single-cloud $\mu$ Sgr template spectrum made it impossible to obtain as good a fit to all of the high-extinction sightlines as the simple two-Gaussian fits described earlier. The same fitting strategy was tried for the 13175\AA\ DIB, and it was found in all cases that the fits obtained using two Gaussians was as good or better than that obtainable using a superposition of up to three $\mu$ Sgr template profiles.

It is conceivable that an acceptable fit to the blue edge of the 11797.5\AA\ DIB profile might be achieved by the addition of a series of closely-spaced extra $\mu$ Sgr template-based velocity components. However, this would require that all of their parameters be fine-tuned in such a way as to build up the observed blue edge. This seems unlikely to occur by chance for multiple sightlines. Furthermore, there is no evidence for such additional clouds in the blue wing of the (stronger) 13175\AA\ DIB.

Our best approximation to a single cloud spectrum does not, when combined with what we know of the cloud velocity structure in front of the sample sources, provide good fits to the observed spectra. This does not appear to be a simple problem of signal-to-noise in the template: there is evidence of blue wings in various sources at 11797.5\AA\ and the profile of that band varies significantly more across our sample than does that of 13175\AA.

Finally, in Figure 7 we plot a parameter (EW/Central Depth) as representative of band width. if velocity structure were a driver of band width (and recall that the strengths of the two bands correlate very well against each other) then a correlation between the two bands might be expected, but none is seen within the uncertainties.

All of the above favors intrinsic variation in the band profiles over velocity structure.

Consequently, we review here some possible profile asymmetry mechanisms in the context of our data:

\begin{enumerate}
\item {\it The blending of multiple overlapping DIBs.} Given that additional DIBs have now been found in the NIR, the possibility of such additional DIB spectrum complexity cannot be wholly discounted. However, the overall similarity of the asymmetry in {\it both} the bands in the high s/n spectra presented here (e.g. Cyg OB2 No. 12 and StRS 354) and in some visual DIBs suggests that this possibility is less likely. We also note that earlier observations of these bands toward other sightlines have given no indication of the presence of nearby additional DIBs.
\item {\it Various anomalous intramolecular broadening processes}. Some of those considered by (e.g.) \citet{walker2001} included predissociation, preionization, and radiationless internal conversion. These were all rejected by them on the grounds that their observed DIB profiles were close to Gaussian, and were fitted comparatively poorly by the Lorentzian profiles expected to arise from such processes. Given the superiority of the Gaussian fits shown in Figure 1 to those using an assumed template spectrum, this also appears to be true of the spectra presented here.
\item {\it Radiative excitation of polar molecules.} More recently, \citet{dahlstrom2013} and \citet{oka2013} discussed extended red wings in visual DIB profiles in the context of the anomalously broadened features seen toward Herschel 36. Based on the modeling of their observational data with linear molecular carrier species, they proposed that such red-winged DIBs may be attributable to polar molecules that are susceptible to radiative excitation. In this scenario, DIBs lacking such a red extension would then most likely be due to non-polar species. \citet{dahlstrom2013} noted that for some of the (non-Her 36) sightlines known to exhibit red-winged DIBs, there are indications of enhanced local ultraviolet radiation fields, high $R(V)$ extinction behavior, and/or abundant dust. It is unlikely that a significant fraction of the diffuse-ISM lines of sight studied here would have a high {\it R(V)} value, but the relatively subtle variations in the red wings of the 11979.5 and 13175\AA\ bands could, in this picture, be a response to local variations in the UV field along the lines of sight. If so, then the apparent independence of the behavior of the red wing in the two bands would indicate that the carriers respond differently to such variations. 

\end{enumerate}

We therefore conclude that the profile differences observed between the sightlines are real, and may be attributable to some combination of effects relating to the formation of the bands themselves.

\subsection{Regional and Intra-regional Variations}

As noted in Section 2, the band strengths vary considerably both between regions and within a region. Specifically, in Cygnus, both StRS 368 and StRS 344 exhibit weak IR DIBs for their extinctions (Figure 4). StRS 368 is also identified with G79.29+0.46, a Luminous Blue Variable (LBV) candidate. This star exhibits local nebulosity seen in both the radio and mid-IR that has been attributed to circumstellar material ejected during earlier phases of greater LBV activity (\citealt{rizzo2014}, \citealt{palau2014}, \citealt{umana2011}). This circumstellar material may be responsible for some fraction of the extinction seen toward this object, although the angular size of the nebulosity suggests it is unlikely to be responsible for $\sim 8^{m}$ of additional extinction, the approximate horizontal displacement of this object from the general trend in Figure 4. Even allowing for the relative complexity of this sightline, we conclude that the NIR DIBs are comparatively weaker toward StRS 368. StRS 344 does not obviously exhibit such circumstellar structure, but is spectroscopically very similar to Cyg OB2 No. 12 \citep{rawlings2000}, which itself may be an LBV (e.g. \citealt{chentsov2013} and references therein). Note also that visible DIBs towards other Cygnus OB2 sources vary considerably in strength per unit extinction (for both {\it A(V)} and {\it E(B-V)}), and the NIR DIBs appear to share this property.

Although the foreground interstellar cloud structures for both Cyg OB2 No. 12 and $\mu$ Sgr have previously been studied using optical high-resolution spectroscopy (e.g. the maximum velocity spread in Na I components of 25 km/s for Cyg OB2 No. 12 reported by \citealt{klochkova2004}), there are few previous observations for most of the StRS objects discussed here. Notable exceptions include StRS 368 (discussed above) and StRS 174. StRS 174 has been identified as "Kleinmann's Anonymous star" or CEN 1 at the center of a young embedded cluster in M17. No H$_3^+$ spectra are available for this sightline. However, this source is already comparatively well-studied (see e.g. \citealt{povich2009} and references therein). \citet{chini1980} and \citet{greve2010} reported {\it A(V)} values of 9.2 and 10.4 respectively, which are broadly consistent with the \citet{rawlings2003} value in Table 1, although \citet{greve2010} reported an {\it R(V)} value of 4.1 -- 4.5, indicating some local deviation from classic diffuse medium extinction. \citet{hoffmeister2008} identified this source as a pair of O4 stars with {\it A(V)} values of 10.2 and 13.5, and that these components themselves were spectroscopic binaries. CO observations by \citet{povich2009} revealed the presence of an extended kinematic shell with a local velocity of 19 km/s associated with star formation, and it seems likely that some of the foreground DIB carriers may be associated with it.

\subsection{Comparison of the NIR DIBs with Other Diffuse Medium Features}

The left and center plots of Figure 8 compare the two observed NIR DIB strengths with those of the two strong, narrow visual DIBs centered near 6284\AA\ and 6614\AA. We have chosen to focus on these two DIBs from \citet{rawlings2003} as they are narrow, strong bands with well-measured equivalent widths (no upper limits) in most/all of the sightlines also observed here at 11979.5 and 13175\AA. No strong correlation is discernible, although a weak correlation with $\lambda$6614 may be present. The large horizontal error bars in these plots reflect the poorer quality of the visual-band data in comparison to those of the IR. The right-hand panels of Figure 8 show the NIR DIB strengths plotted against the optical depth of the 3.4$\mu$m short-chained aliphatic hydrocarbon dust absorption feature. Although there are few sources for which all three features are sampled, the dataset indicates that the carriers of the IR DIBs and the solid-phase 3.4$\mu$m vary independently of each other.

 \subsection{Constraints on possible DIB Carriers}

We have confirmed that the features at 11797.5\AA~and 13175\AA~ are DIBs and exhibit intrinsic profile structure unrelated to velocity. Unlike the {\it H}-band DIBs reported by \citet{geballe2011}, these DIBs are attributable to electronic transitions, as vibrational overtones would be very weak in comparison.  The existence of these low frequency DIBs indicates the presence of one or two related carriers that are able to exhibit very low-lying electronically-excited states.  This might suggest that the molecules involved are radicals, exhibiting open electron shells. Their diffuse ISM ubiquity and comparative scarcity inside dense molecular clouds supports the role of some form of photoprocessing in their production. As with DIBs seen at other wavelengths, the carriers are unlikely to be small, stable molecules, as these always tend to have their lowest transitions in the ultraviolet, and these cases are more extreme.
  
\section{CONCLUSIONS AND SUMMARY}

We have measured the strengths and profiles of two spectral absorption bands near 11797.5\AA~and 13175\AA~that had been previously identified as DIBs, and support this identification. The 13175\AA~DIB is always found to be markedly stronger than the 11797.5\AA~DIB, presenting implications for the f values and/or the relative abundances of the DIB carriers. This is entirely consistent with earlier observations of these DIBs toward other sightlines. The equivalent widths of these two bands correlate very well across a sample which covers a wide range of extinction and Galactic location. This suggests that the carriers are one or two related species that are able to exhibit very low-lying electronically-excited states. Two high-extinction sources exhibit weak DIB outlier behavior in Figure 4, StRS 344 and StRS 368. Some of the visual extinction towards StRS 368 may be attributable to additional extinction arising from circumstellar nebulosity, but its DIBs are still thought to be relatively weak. Per unit extinction, there is the suggestion of a correlation between the 13175\AA~and 11797.5\AA~DIB strengths, which is statistically skewed by the anomalously strong bands measured toward $\mu$ Sgr. Within the rest of the sample, the ratioed strengths are fairly clustered.

We find that the profiles of the two NIR DIBs observed can be fitted well by a superposition of two Gaussians. The second Gaussian is often necessary in order to produce a good fit to a red wing. Furthermore, we find that for all of the high-extinction sightlines observed, profile fitting performed using two Gaussians produces fits as good or better than fits using DIB template profiles derived from our $\mu$ Sgr "single cloud" sightline.  We discount the possibility of a simple blending of multiple features, and conclude that the DIB profile variations seen between sightlines are at least partially intrinsic, and must be attributed to the band formation process rather than to velocity structure. We find that the observed profiles of both DIBs can be divided into similar groups, as shown in Figures 2 and 3.

Neither of the DIBs observed exhibits complex substructure at the spectral resolutions presented here. However, the substructure previously observed in some of the narrow visual wavelength DIBs \citep{sarre1995} was extremely narrow, and may suggest that higher-resolution spectroscopy is needed in order to detect comparable substructure in the NIR DIBs. At the resolution of our observations, the profiles have been shown to be similar to those of visual-band DIBs.

We find no strong evidence of correlations between the NIR DIBs and the two strong, narrow visual DIBs near 6284 and 6614\AA, suggesting that they are attributable to different carrier species. We find that the observed NIR DIB strengths also vary independently of the $\tau$(3.4$\mu$m) that traces short-chained aliphatic hydrocarbons in the diffuse ISM.



\acknowledgments

The authors wish to thank Ben McCall for his input on early versions of this paper. MGR acknowledges support from the National Radio Astronomy Observatory. The National Radio Astronomy Observatory is a facility of the National Science Foundation operated under cooperative agreement by Associated Universities, Inc. The United Kingdom Infrared Telescope was operated by the Joint Astronomy Centre on behalf of the Science and Technology Facilities Council of the U.K. This research has made use of NASA's Astrophysics Data System.



{\it Facilities:} \facility{UKIRT}.




\clearpage



\renewcommand{\thefigure}{\arabic{figure}\alph{subfigure}}
\setcounter{subfigure}{1}

\begin{figure}
\begin{center}
\vspace{-1.0cm}
\hspace{-2.0cm}
\includegraphics[angle=0,scale=0.85]{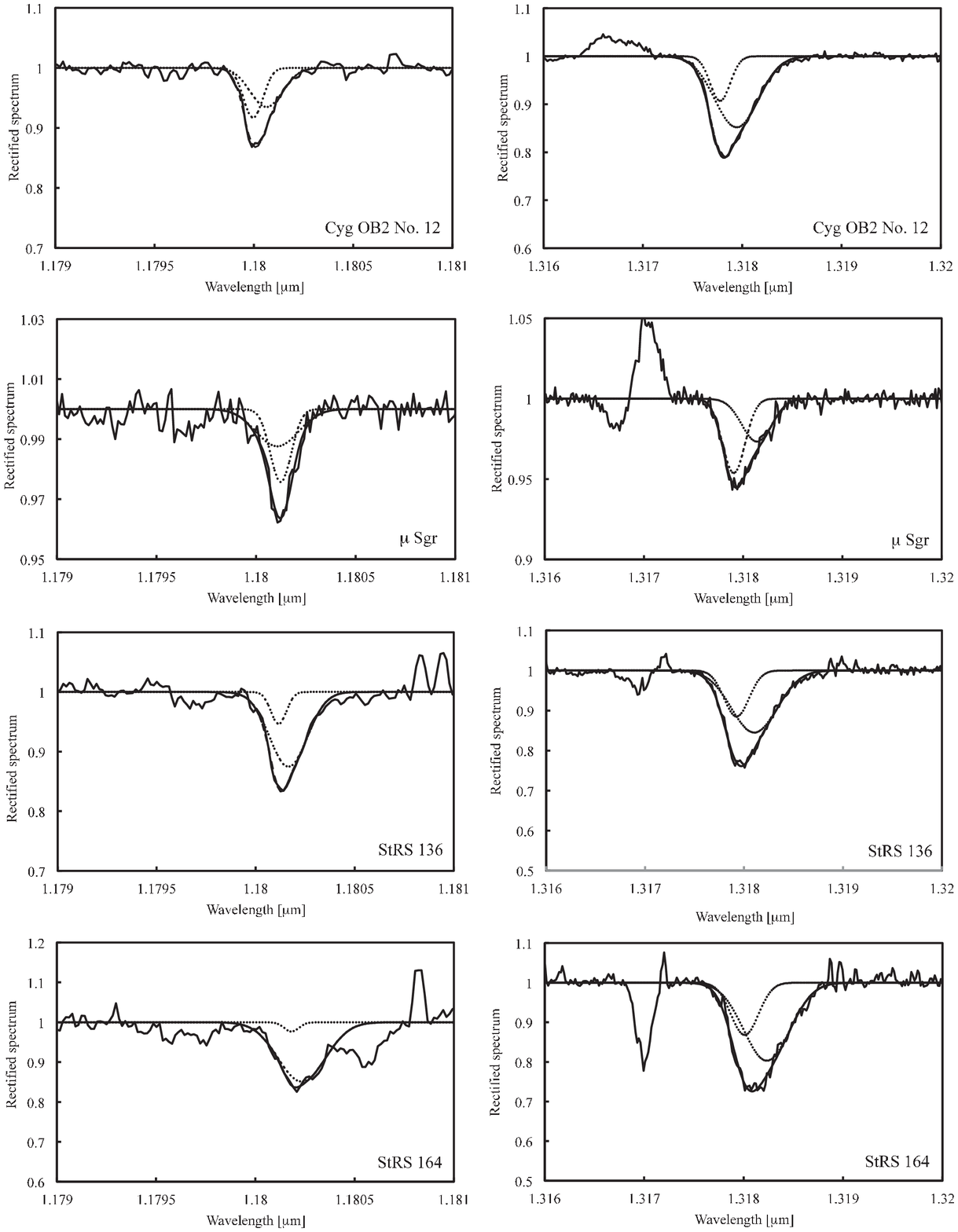}
\vspace{-0.5cm}
\caption{Echelle spectra of the targets listed in Table~\ref{tbl-1}. The left-hand panels show the spectra centered around 11797.5\AA\ and the right-hand panels show the spectral centered around 13175\AA. Dotted lines denote the Gaussian fitting components. The sources names are indicated in the individual panels.\label{fig1a}}
\end{center}
\end{figure}

\addtocounter{figure}{-1}
\addtocounter{subfigure}{1}

\begin{figure}
\begin{center}
\vspace{-1.0cm}
\hspace{-2.0cm}
\includegraphics[angle=0,scale=0.85]{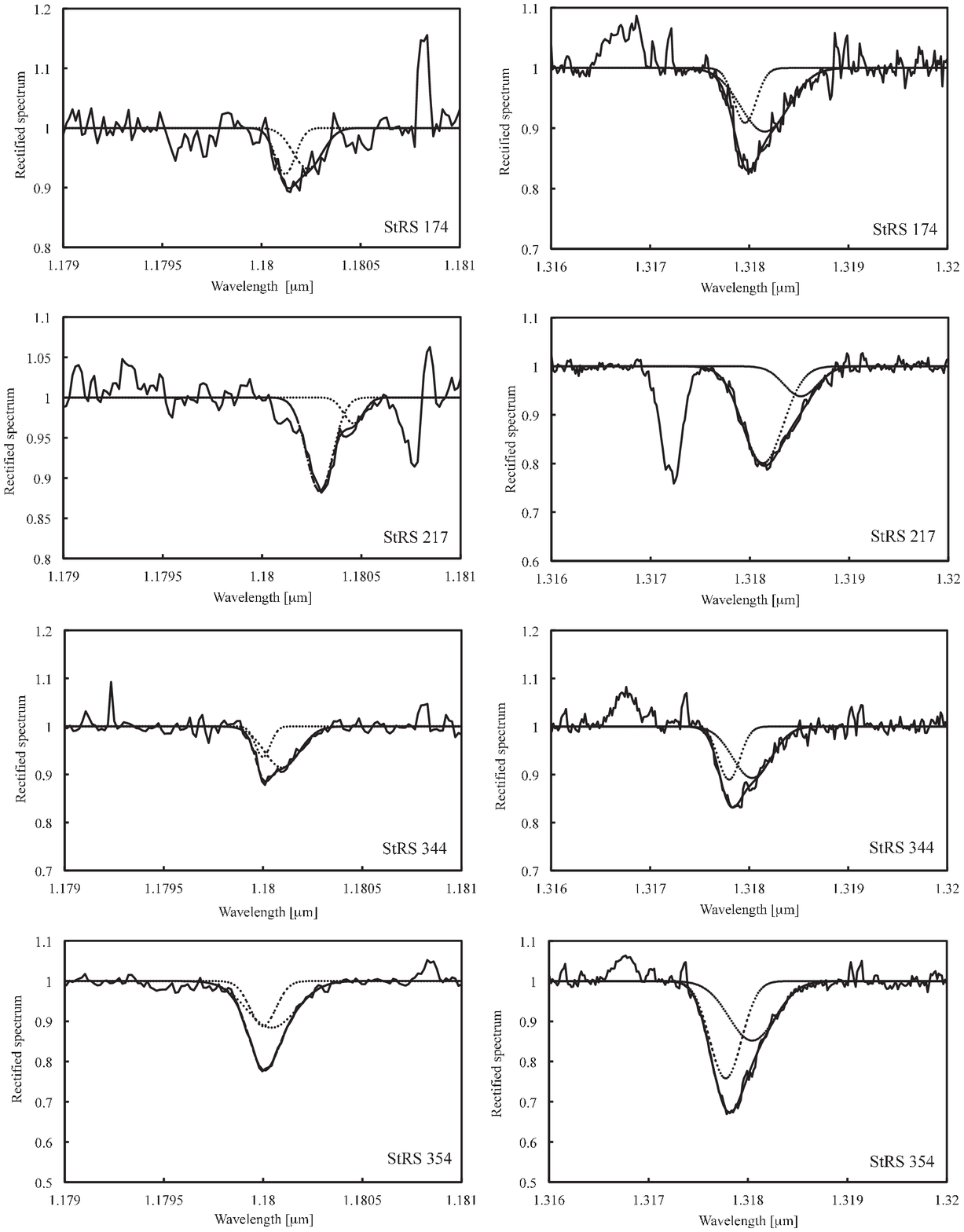}
\vspace{-0.5cm}
\caption{Echelle spectra of the targets listed in Table~\ref{tbl-1} (continued). The left-hand panels show the spectra centered around 11797.5\AA\ and the right-hand panels show the spectral centered around 13175\AA. Dotted lines denote the Gaussian fitting components. The sources names are indicated in the individual panels.\label{fig1b}}
\end{center}
\end{figure}

\addtocounter{figure}{-1}
\addtocounter{subfigure}{1}

\begin{figure}
\begin{center}
\vspace{-6.0cm}
\hspace{-2.0cm}
\includegraphics[angle=0,scale=0.85]{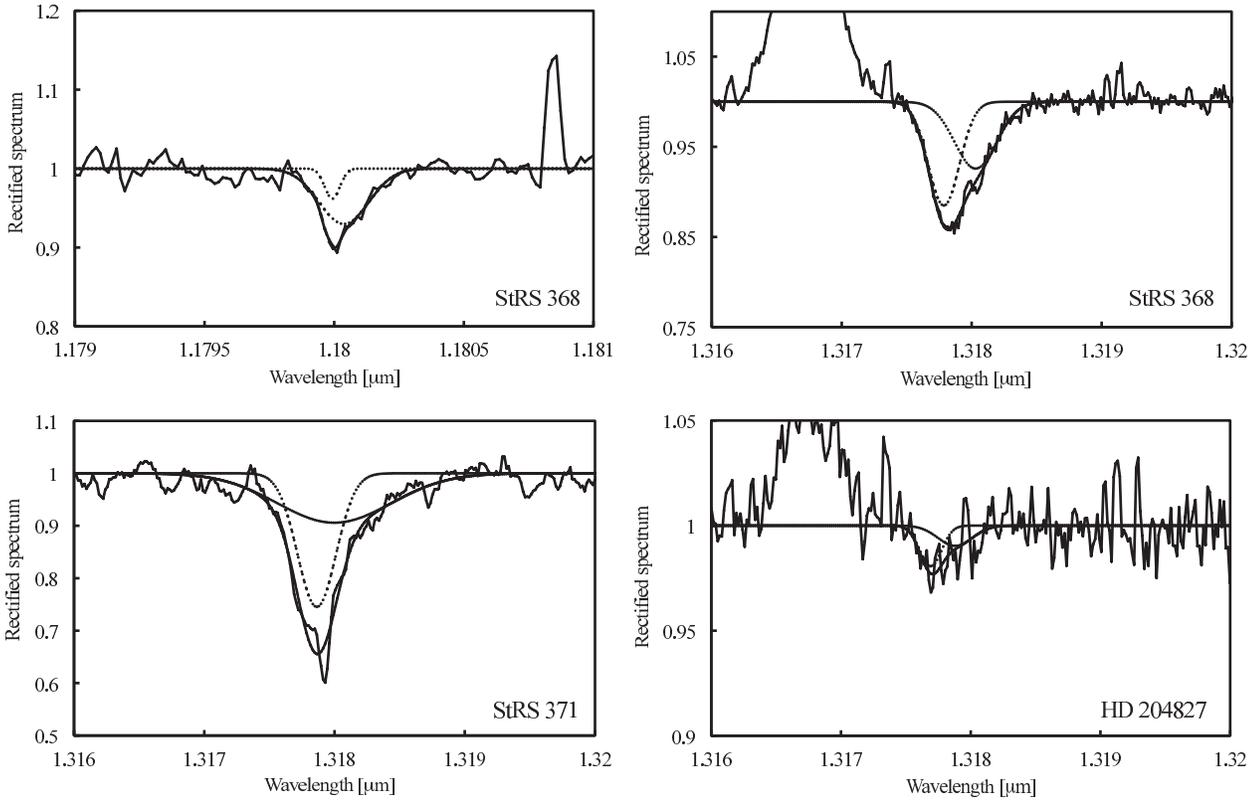}
\caption{Echelle spectra of the targets listed in Table~\ref{tbl-1} (continued). The upper left-hand panel shows the StRS 368 spectrum centered around 11797.5\AA\ and the right-hand panels show the spectral centered around 13175\AA\ for the same source. The two lower panels show spectra of sources for 13175\AA\ for the two sources for which no 11797.5\AA\ spectra were obtained. Dotted lines denote the Gaussian fitting components. The sources names are indicated in the individual panels.\label{fig1c}}
\end{center}
\end{figure}

\renewcommand{\thefigure}{\arabic{figure}}

\clearpage


\begin{figure}
\begin{center}
\vspace{-0.5cm}
\includegraphics[angle=0,scale=0.85]{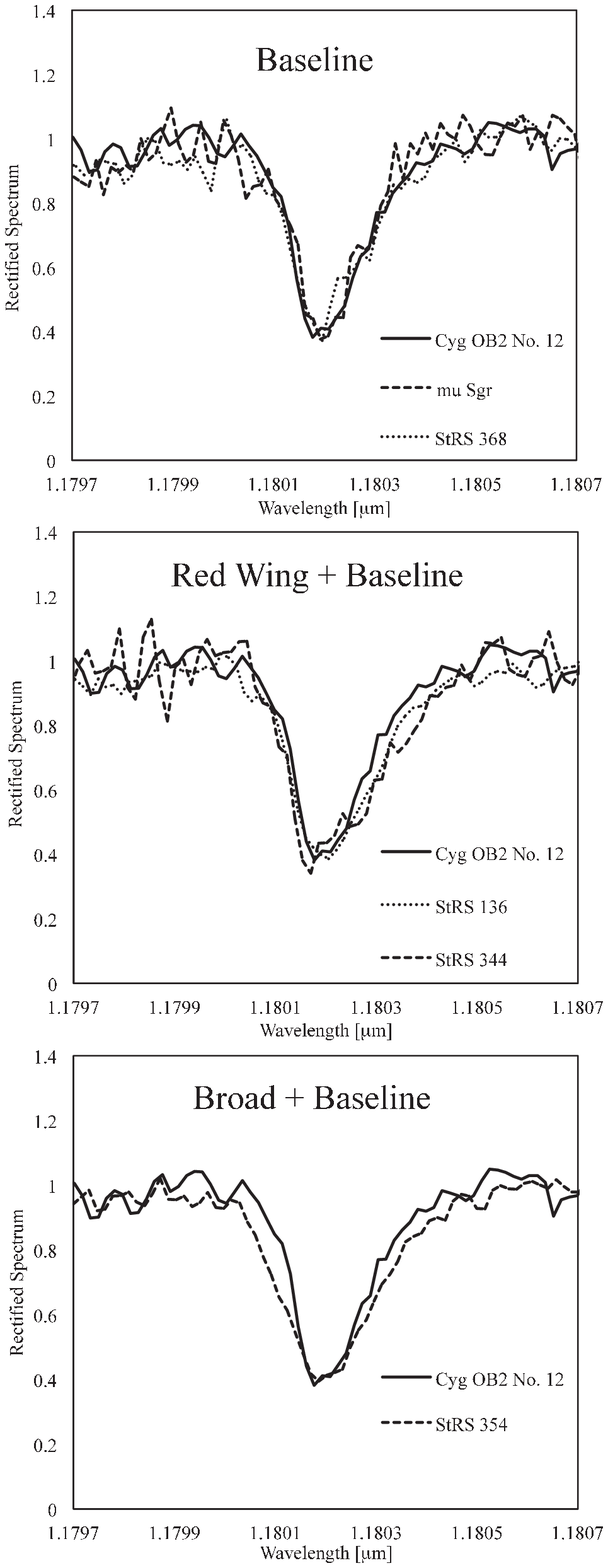}
\vspace{-0.3cm}
\caption{Classification of the different 11797.5\AA\ profiles. The profiles have been linearly rescaled vertically and the central wavelengths shifted for direct comparison. Cyg OB2 No. 12 is used as a representative baseline reference profile in all three panels.\label{fig2}}
\end{center}
\end{figure}

\begin{figure}
\begin{center}
\vspace{-0.5cm}
\includegraphics[angle=0,scale=0.8]{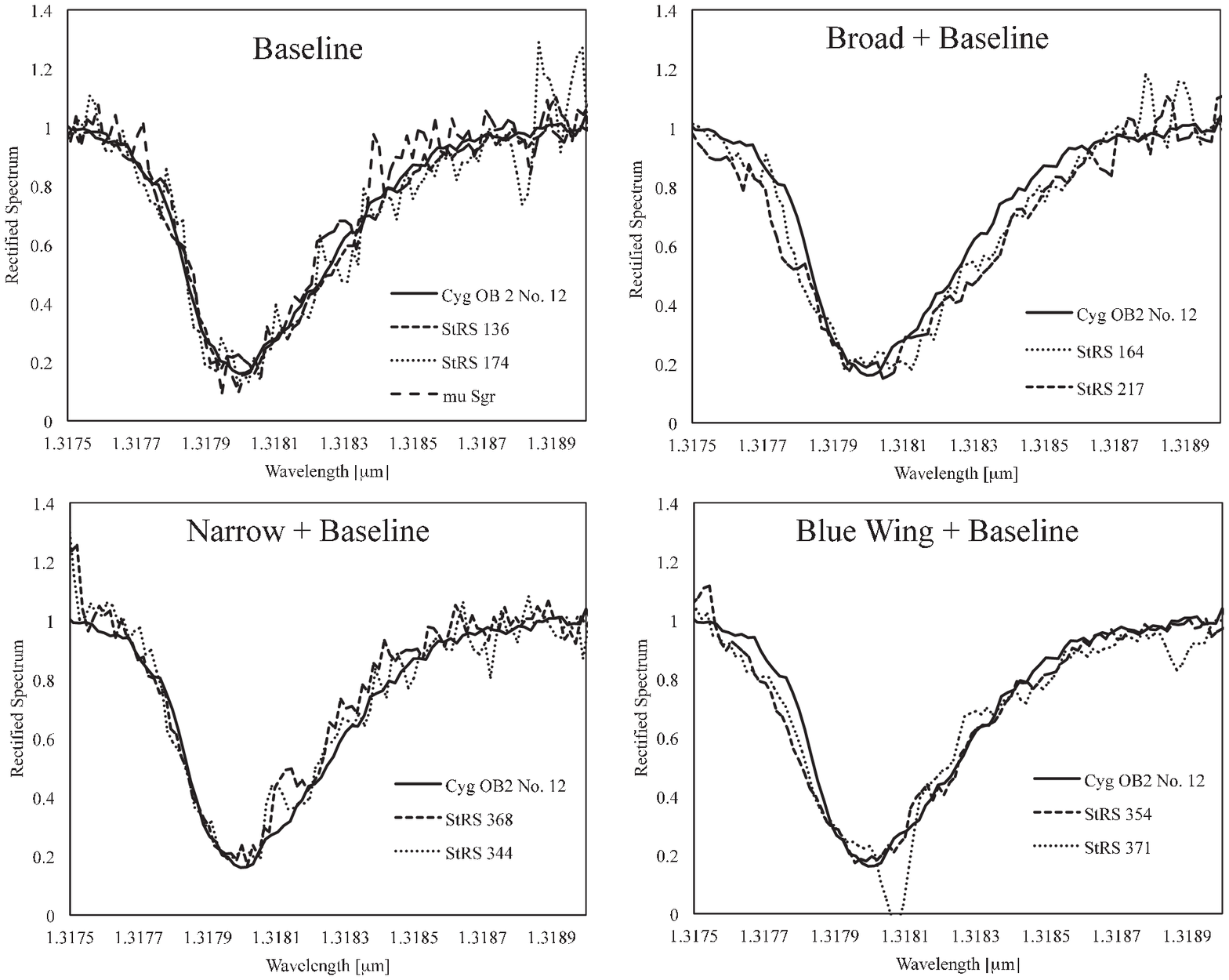}
\caption{Classification of the different 13175\AA\ profiles. The profiles have been linearly rescaled vertically and the central wavelengths shifted for direct comparison. Cyg OB2 No. 12 is used as a representative baseline reference profile in all four panels.\label{fig3}}
\end{center}
\end{figure}

\begin{figure}
\begin{center}
\vspace{-0.8cm}
\includegraphics[angle=0,scale=0.70]{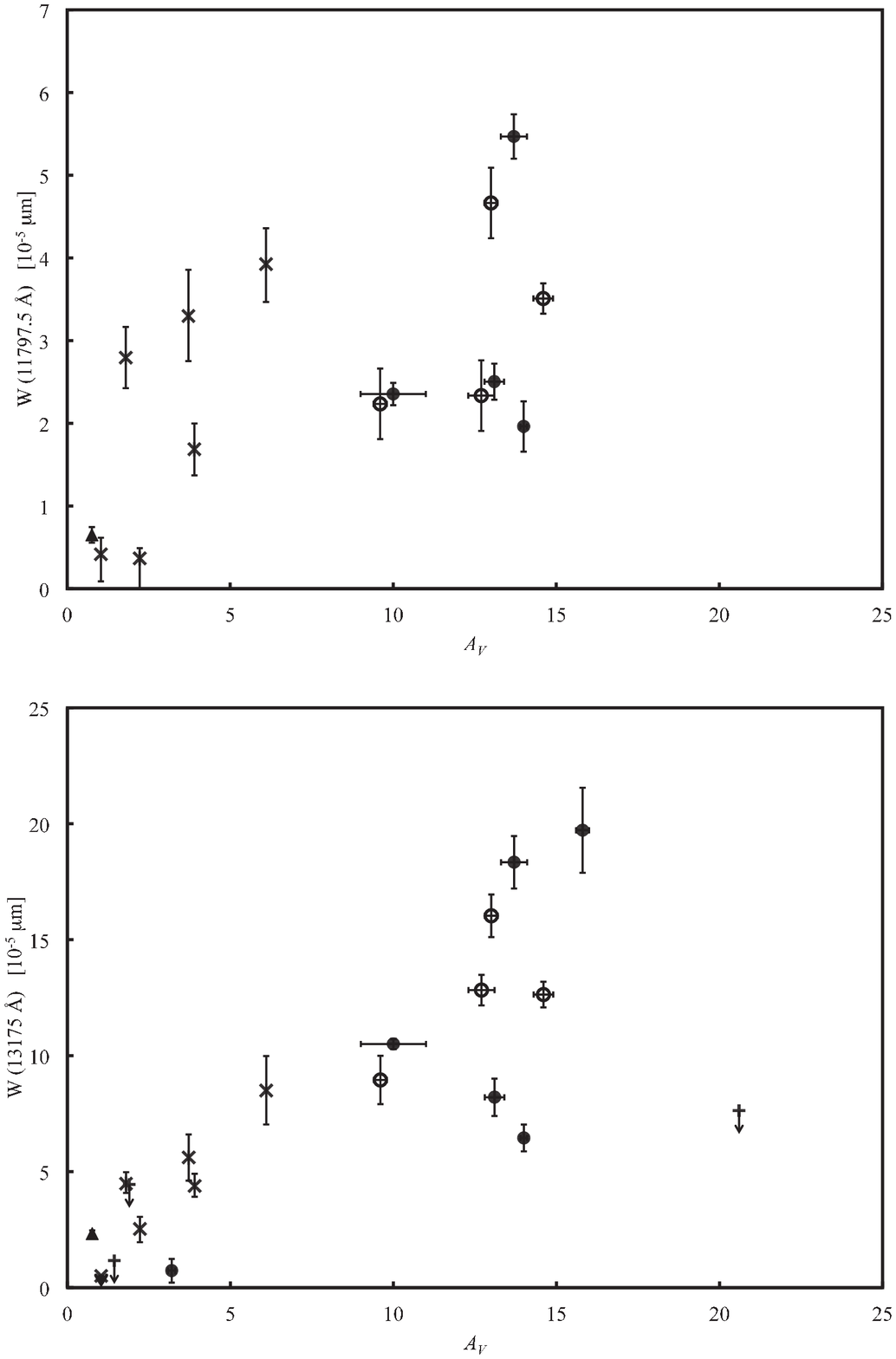}
\vspace{-0.8cm}
\caption{Equivalent width versus A(V) for the two observed NIR DIBs. The upper panel shows the 11797.5\AA~DIB equivalent widths versus A(V). The lower panel shows the 13175\AA~DIB equivalent widths versus A(V). In both panels, filled triangles represent $\mu$ Sgr, filled circles represent sources with high Galactic longitudes (nearer Cygnus) and open circles represent sources with low Galactic longitudes (closer to the Galactic Centre). The error bars reflect the root mean square noise on the spectra. The "$\times$" symbols represent the data of \cite{joblin1990} and in the lower panel, the "$+$" symbols the upper limits obtained by \citet{adamson1994}.\label{fig4}}
\end{center}
\end{figure}

\begin{figure}
\begin{center}
\vspace{-0.5cm}
\includegraphics[angle=0,scale=0.75]{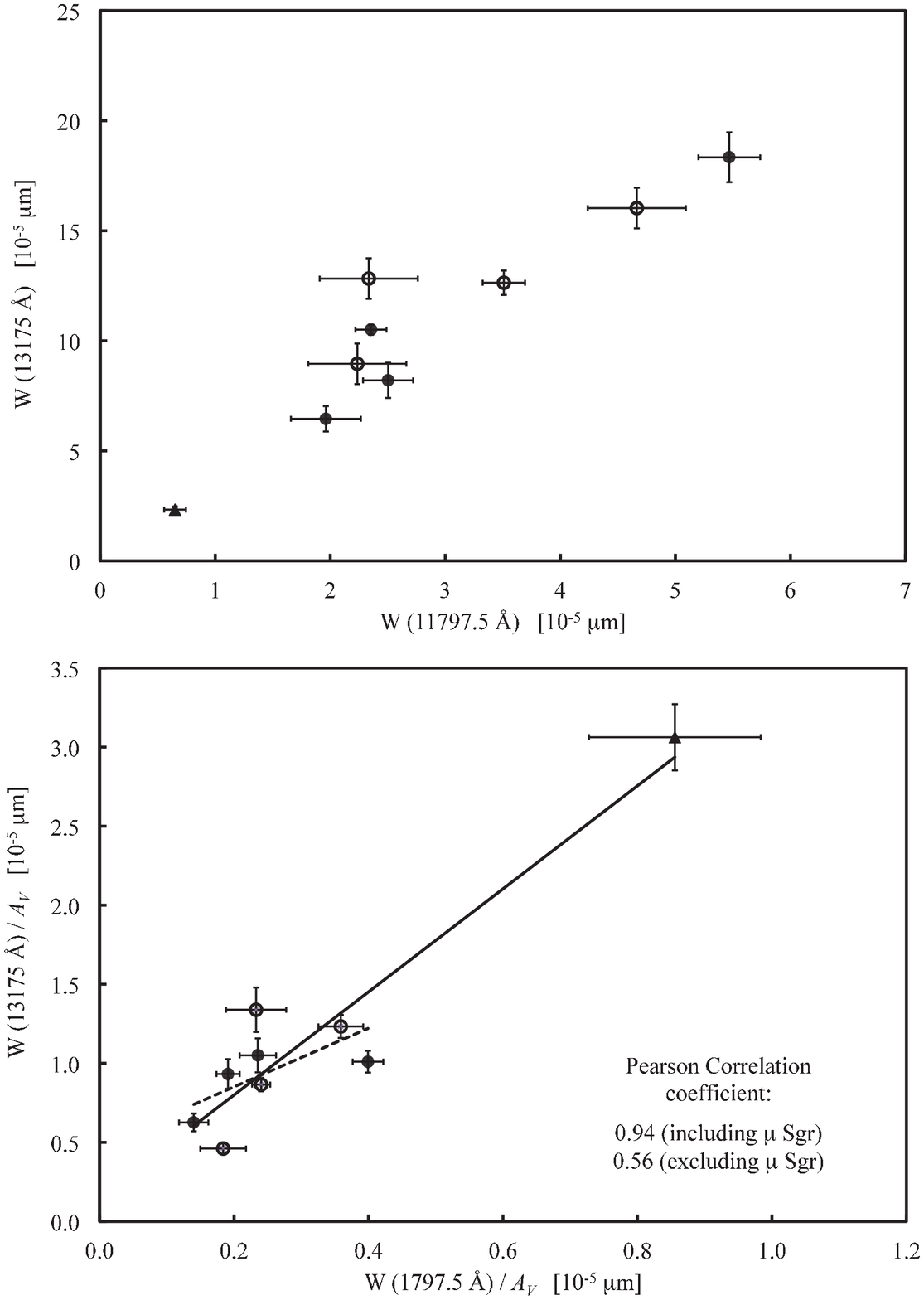}
\caption{The strengths of the two DIBs, as measured (upper panel) and per unit extinction (lower panel). In both panels, filled triangles represent $\mu$ Sgr, filled circles represent sources with high Galactic longitudes (nearer Cygnus) and open circles represent sources with low Galactic longitudes (closer to the Galactic Centre). The error bars reflect the root mean square noise on the spectra.\label{fig5}}
\end{center}
\end{figure}

\begin{figure}
\begin{center}
\vspace{-0.5cm}
\includegraphics[angle=0,scale=0.65]{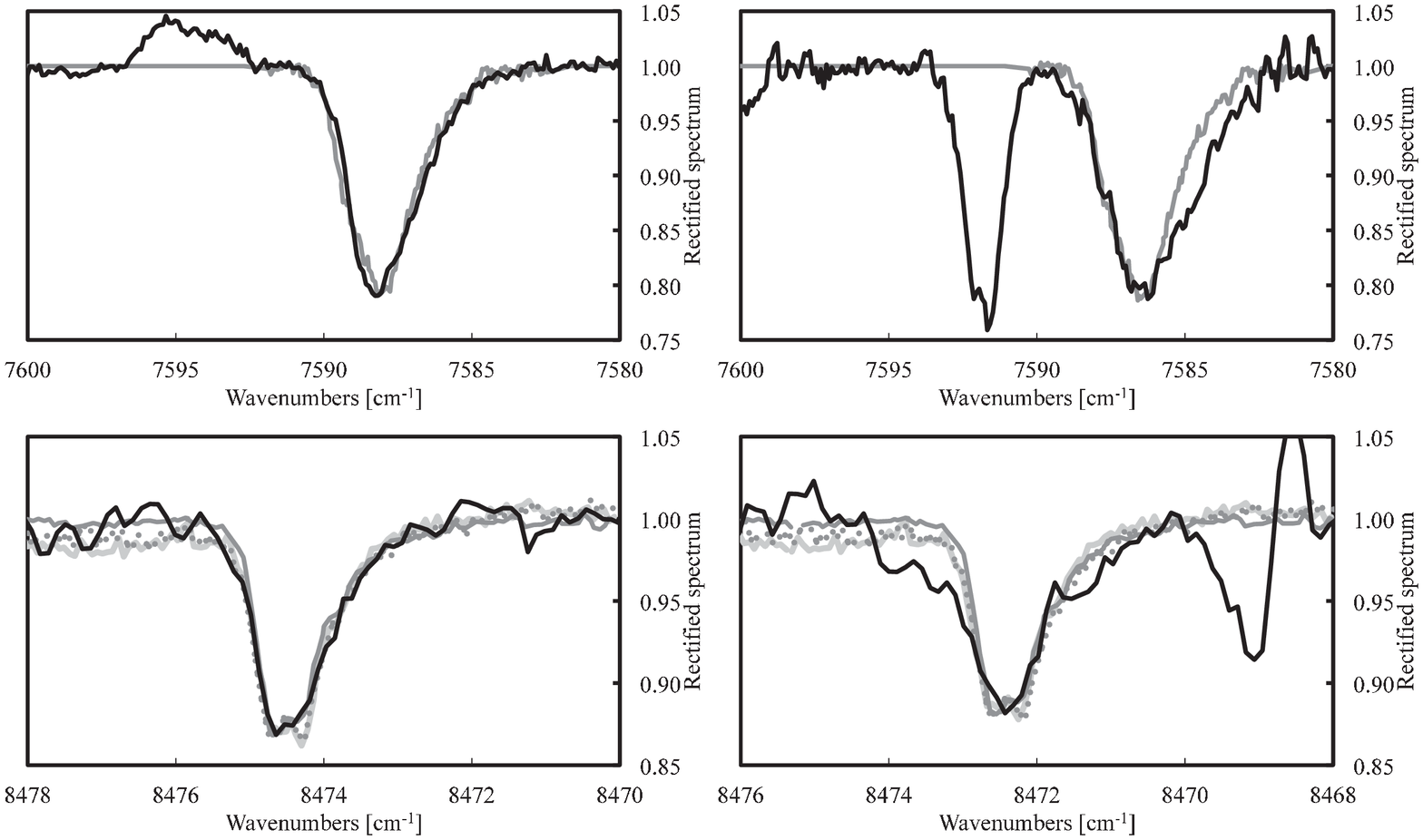}
\vspace{-0.5cm}
\caption{A comparison of selected observed NIR DIB profiles against those of previously observed average visual DIBs. The upper panels show the observed 13175\AA\ DIB profiles (black solid lines) toward Cyg OB2 No. 12 (left) and StRS 217 (right). The dark gray lines in these plots show the averaged ``zeta" profile spectrum of the 5780\AA\ DIB from Figure 8 of \citet{krelowski1997}, rescaled to the same peak central wave number, and linearly rescaled for depth. The lower panels show the observed 11797.5\AA\ DIB profiles (black lines) toward the same two sightlines. The three average profiles of the 5797\AA\ DIB from Figure 9 of \citet{krelowski1997} are shown in the lower panels, again after matching central peak and rescaling for band depth. Their  solid dark gray, solid light gray and dotted dark gray lines represent their ``zeta", ``sigma" and ``rho" profiles respectively. Only the ``zeta" spectrum of \citet{krelowski1997} is included in the upper two panels, as this profile was reported to be invariant between the cloud types.\label{fig6}}
\end{center}
\end{figure}

\begin{figure}
\begin{center}
\vspace{-0.5cm}
\includegraphics[angle=0,scale=0.65]{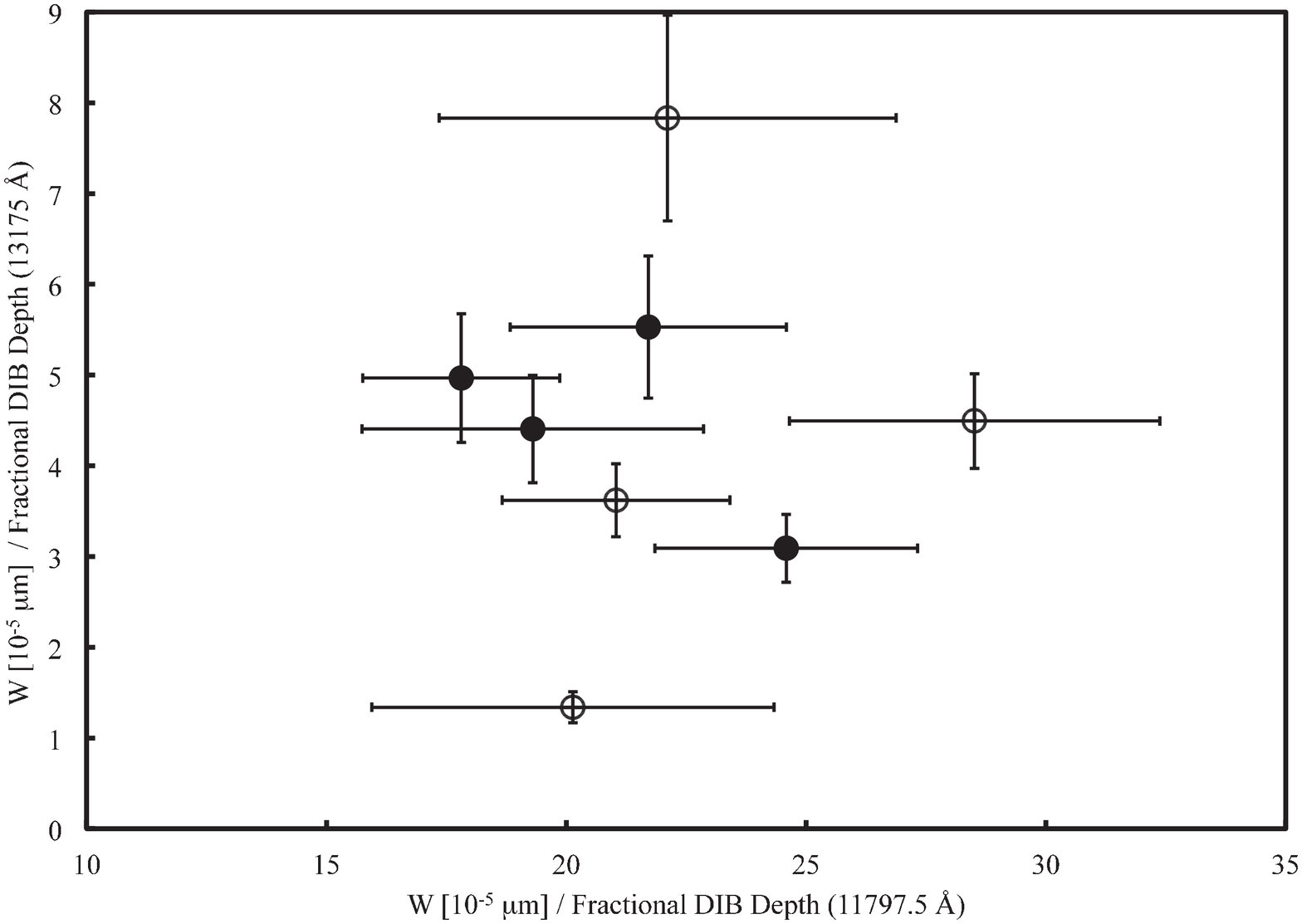}
\caption{The widths of the two observed NIR DIBs plotted against each other. The $\mu$ Sgr data point (with W/depth values of $\sim$20$\times 10^{-5}\mu$m and $\sim$55$\times 10^{-5}\mu$m for the 11797.5\AA\ and 13175\AA\ DIBs respectively) has been omitted for clarity. Filled circles represent sources with high Galactic longitudes (nearer Cygnus) and open circles represent sources with low Galactic longitudes (closer to the Galactic Centre).\label{fig7}}
\end{center}
\end{figure}

\begin{figure}
\begin{center}
\vspace{-0.5cm}
\includegraphics[angle=0,scale=0.72]{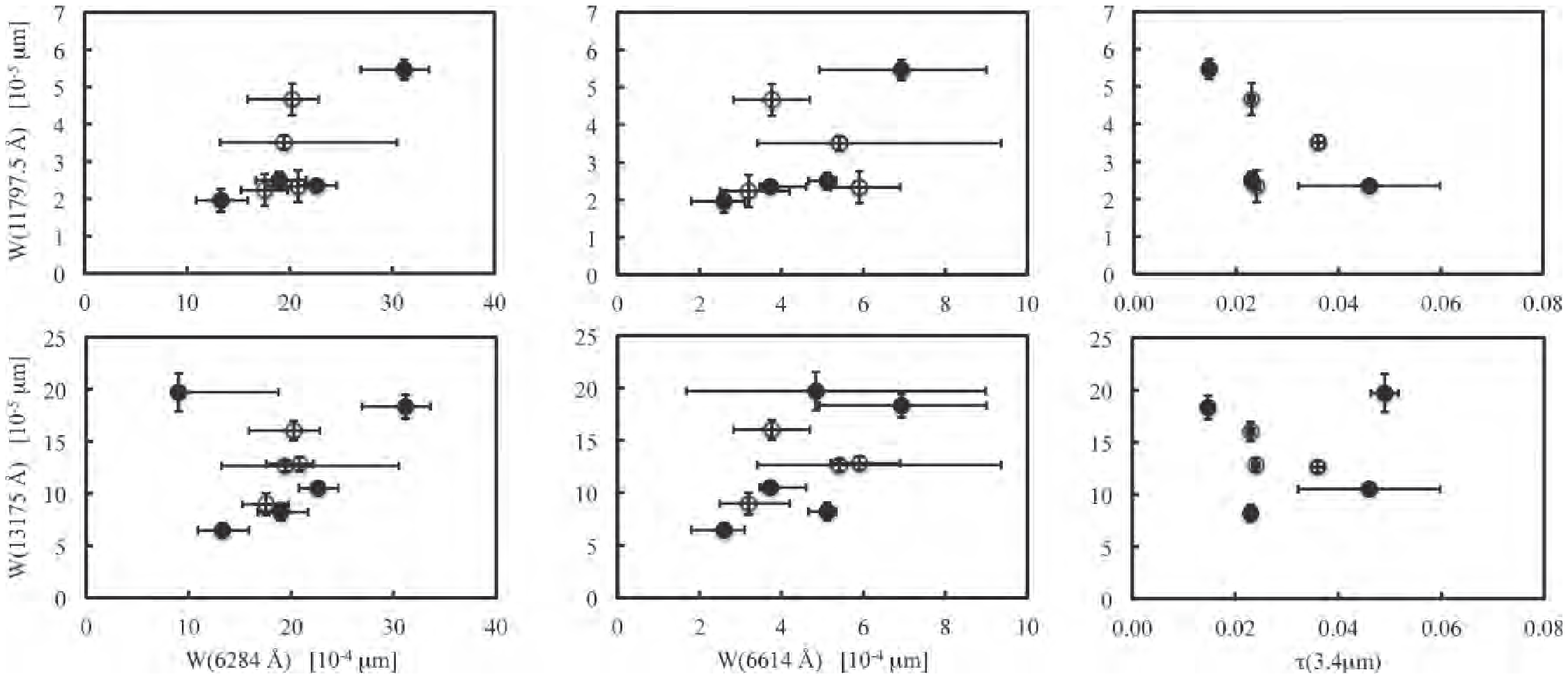}
\caption{The strengths of the two NIR DIBs plotted against the narrow visual DIBs centered near 6284 and 6614\AA\ (left and center column plots), and against the optical depth measurements of the 3.4$\mu$m short-chained aliphatic hydrocarbon dust absorption feature (right column plots). Filled circles represent sources with high Galactic longitudes (nearer Cygnus) and open circles represent sources with low Galactic longitudes (closer to the Galactic Centre).\label{fig8}}
\end{center}
\end{figure}







\clearpage

\begin{deluxetable}{cccccc}
\tabletypesize{\scriptsize}
\tablecaption{List of observed science targets.\label{tbl-1}}
\tablewidth{0pt}
\tablehead{
\colhead{Source} & \colhead{$\alpha$} & \colhead{$\delta$} & \colhead{Type} & \colhead{{\it A(V)}} & \colhead{Note} \\
 & \multicolumn{2}{c}{(J2000.0)} & & & }
\startdata
StRS 136 & 17\textsuperscript{h}47\textsuperscript{m}56\textsuperscript{s}.08 & $-29^\circ11'43''.9$ & B8 -- A9I$\tablenotemark{a}$ & $14.6\pm0.3\tablenotemark{a}$ & \\
StRS 164 & 18\textsuperscript{h}16\textsuperscript{m}18\textsuperscript{s}.75 & $-16^\circ35'46''.8$ & B8 -- A9I$\tablenotemark{a}$ & $13.0\pm0.2\tablenotemark{a}$ & \\
StRS 174 & 18\textsuperscript{h}20\textsuperscript{m}29\textsuperscript{s}.86 & $-16^\circ10'44''.9$ & B5 -- B9.5$\tablenotemark{a}$ & $9.6\pm0.2\tablenotemark{a}$ & \\
StRS 217 & 18\textsuperscript{h}32\textsuperscript{m}58\textsuperscript{s}.12 & $-08^\circ46'17''.0$ & B8 -- A9I$\tablenotemark{a}$ & $12.7\pm0.4\tablenotemark{a}$ & \\
StRS 344 & 20\textsuperscript{h}20\textsuperscript{m}37\textsuperscript{s}.86 & $+41^\circ09'47''.7$ & O8 -- B7$\tablenotemark{a}$ & $13.1\pm0.3\tablenotemark{a}$ & \\
StRS 354 & 20\textsuperscript{h}29\textsuperscript{m}15\textsuperscript{s}.36 & $+37^\circ51'01''.9$ & O7 -- B3$\tablenotemark{a}$ & $13.7\pm0.4\tablenotemark{a}$ & \\
StRS 368 & 20\textsuperscript{h}31\textsuperscript{m}42\textsuperscript{s}.28 & $+40^\circ21'59''.1$ & B6 -- B9.5$\tablenotemark{a}$ & $14.0\pm0.1\tablenotemark{a}$ & Expanding shell \\
StRS 371 & 20\textsuperscript{h}33\textsuperscript{m}23\textsuperscript{s}.91 & $+40^\circ36'44''.2$ & Early$\tablenotemark{a}$ & $15.8\pm0.2\tablenotemark{a}$ & Possible binary \\
Cyg OB2\#12 & 20\textsuperscript{h}32\textsuperscript{m}40\textsuperscript{s}.96 & $+41^\circ14'29''.1$ & B5Ia$^+$~$\tablenotemark{a}$ & $10.0\pm1\tablenotemark{a}$ & \\
HD 204827 & 21\textsuperscript{h}28\textsuperscript{m}57\textsuperscript{s}.761 & $+58^\circ44'23''.24$ & B0V\tablenotemark{b} & $3.4\pm0.1\tablenotemark{b}$ & Spectroscopic binary\\
$\mu$ Sgr & 18\textsuperscript{h}13\textsuperscript{m}45\textsuperscript{s}.81 & $-21^\circ03'31''.8$ & B2III & $0.76\pm0.03$ & ``Single cloud'' sightline;\\
 & & & & & Eclipsing binary \\
\enddata
\tablenotetext{a}{\citet{rawlings2003}.}
\tablenotetext{b}{\citet{hobbs2008}.}
\end{deluxetable}


\clearpage

\clearpage

\begin{deluxetable}{ccccccccccccc}
\tabletypesize{\scriptsize}
\rotate
\tablecaption{List of characteristics derived from fitting the normalized $\lambda$11797.5 spectra with two Gaussians. }\label{tbl-2}
\tablewidth{0pt}
\tablehead{
\colhead{Source} &  \multicolumn{5}{c}{Gaussian 1} & \multicolumn{5}{c}{Gaussian 2} & & \colhead{Equivalent Width, W} \\
 & \colhead{$\lambda$} & \colhead{$\sigma$} & \colhead{Depth} & \colhead{FWHM} & \colhead{Area} & & \colhead{$\lambda$} & \colhead{$\sigma$} & \colhead{Depth} & \colhead{FWHM} & \colhead{Area} & (Total Area) \\
 & \colhead{[$\mu$m]} & [$10^{-5} \mu$m] & & [$10^{-5} \mu$m] & [$10^{-5} \mu$m]& & \colhead{[$\mu$m]} & [$10^{-5} \mu$m] & & [$10^{-5} \mu$m] & [$10^{-5} \mu$m] & [$10^{-5} \mu$m] \\
}
\startdata
Cyg OB2 No. 12 & 1.18000 & 5 & 0.08290 & 12 & 1.02127 &  & 1.18006 & 8 & 0.06509 & 19 & 1.33253 & $2.35\pm0.135$ \\
HD 204827 &  \multicolumn{12}{c}{S/N too low for DIB detection} \\
$\mu$ Sgr & 1.18011 & 11 & 0.01241 & 25 & 0.330347 &  & 1.18012 & 5 & 0.02429 & 12 & 0.319911 & $0.650\pm0.0944$ \\
StRS 136\tablenotemark{a} & 1.18017 & 9 & 0.12540 & 22 & 2.9794 &  & 1.18012 & 4 & 0.05357 & 9 & 52.9475 & $3.51\pm0.184$ \\
StRS 164 & 1.18023 & 12 & 0.14873 & 28 & 4.44865 &  & 1.18018 & 4 & 0.02334 & 9 & 0.216047 & $4.66\pm0.427$ \\
StRS 174 & 1.18012 & 5 & 0.07725 & 12 & 0.982127 & & 1.18023 & 7 & 0.06808 & 17 & 1.25276 & $2.23\pm0.462$ \\
StRS 217 &  1.18029 & 7 & 0.11659 & 16 & 1.93522 & & 1.18046 & 5 & 0.03211 & 12 & 0.399495 & $2.33\pm0.348$ \\
StRS 344 & 1.18000 & 4 & 0.06389 & 9 & 0.600356 &  & 1.18010 & 9 & 0.08580 & 21 & 1.90272 & $2.50\pm0.218$ \\
StRS 354 & 1.18000 & 7 & 0.11197 & 15 & 1.82723 &  & 1.18004 & 13 & 0.11602 & 29 & 3.64114 & $5.47\pm0.268$ \\
StRS 368 & 1.17999 & 3 & 0.03818 & 7 & 0.278176 &  & 1.18004 & 10 & 0.07019 & 22 & 1.68322 & $1.96\pm0.304$ \\
StRS 371 &  \multicolumn{12}{c}{No observation made near 1.2$\mu$m} \\
\enddata
\tablenotetext{a}{DIB blended.}
\end{deluxetable}



\clearpage

\begin{deluxetable}{ccccccccccccc}
\tabletypesize{\scriptsize}
\rotate
\tablecaption{List of characteristics derived from fitting the normalized $\lambda$13175 spectra with two Gaussians. }\label{tbl-3}
\tablewidth{0pt}
\tablehead{
\colhead{Source} &  \multicolumn{5}{c}{Gaussian 1} & \multicolumn{5}{c}{Gaussian 2} & & \colhead{Equivalent Width, W} \\
 & \colhead{$\lambda$} & \colhead{$\sigma$} & \colhead{Depth} & \colhead{FWHM} & \colhead{Area} & & \colhead{$\lambda$} & \colhead{$\sigma$} & \colhead{Depth} & \colhead{FWHM} & \colhead{Area} & (Total Area) \\
 & \colhead{[$\mu$m]} & [$10^{-5} \mu$m] & & [$10^{-5} \mu$m] & [$10^{-5} \mu$m]& & \colhead{[$\mu$m]} & [$10^{-5} \mu$m] & & [$10^{-5} \mu$m] & [$10^{-5} \mu$m] & [$10^{-5} \mu$m] \\
}
\startdata
Cyg OB2 No. 12 & 1.31778 & 10 & 0.09302 & 24 & 2.35199 & & 1.31794 & 22 & 0.14821 & 52 & 8.15096 & $10.5\pm0.0218$ \\
HD 204827 & 1.31769 & 8 & 0.01934 & 20 & 0.40715 & & 1.31788 & 13 & 0.00966 & 31 & 0.319662 & $0.727\pm0.511$ \\
$\mu$ Sgr & 1.31790 & 11 & 0.04623 & 27 & 1.31722 & & 1.31814 & 15 & 0.02670 & 35 & 1.01049 & $2.33\pm0.138$ \\
StRS 136 & 1.31792 & 13 & 0.11494 & 30 & 3.62042 & & 1.31810 & 23 & 0.15518 & 54 & 9.01531 & $12.6\pm0.555$ \\
StRS 164 & 1.31801 & 14 & 0.13175 & 33 & 4.57797 & & 1.31823 & 23 & 0.19687 & 55 & 11.4537 & $16.0\pm0.92$ \\
StRS 174 & 1.31796 & 11 & 0.09108 & 26 & 2.5417 & & 1.31816 & 24 & 0.10558 & 57 & 6.41216 & $8.95\pm1.04$ \\
StRS 217 & 1.31813 & 20 & 0.19915 & 48 & 10.1943 & & 1.31851 & 17 & 0.06164 & 40 & 1.63266 & $12.8\pm0.658$ \\
StRS 344 & 1.31780 & 11 & 0.11048 & 25 & 1.99708 & & 1.31803 & 19 & 0.10706 & 46 & 5.20709 & $8.20\pm0.804$ \\
StRS 354 & 1.31778 & 16 & 0.24210 & 38 & 9.68732 & & 1.31804 & 23 & 0.14815 & 55 & 8.65198 & $18.3\pm1.13$ \\
StRS 368 & 1.31779 & 12 & 0.11518 & 28 & 3.45811 & & 1.31803 & 16 & 0.07420 & 38 & 2.99487 & $6.45\pm0.578$ \\
StRS 371\tablenotemark{a} & 1.31786 & 15 & 0.25473 & 35 & 9.43213 & & 1.31800 & 44 & 0.09419 & 102 & 10.2867 & $19.7\pm1.83$ \\
\enddata
\tablenotetext{a}{DIB blended.}
\end{deluxetable}




\end{document}